\documentstyle[12pt,aaspp4]{article}

\begin{document}

%definitions
\def \coto{$^{12}$CO(2--1) \/}
\def \cooz{$^{12}$CO(1--0) \/}
\def \thco{$^{13}$CO(1--0) \/}
\def \twco{$^{12}$CO \/}
\def \vlsr{$V_{LSR}$ \/}
\slugcomment{To appear in {\it The Astrophysical Journal Letters}}

\title{The Mass-Velocity and Position-Velocity Relations in Episodic
Outflows}
\author{H\'ector G. Arce \& Alyssa A. Goodman}
\affil{Harvard--Smithsonian Center for Astrophysics, 60 Garden St.,
Cambridge, MA 02138}
\authoremail{harce@cfa.harvard.edu, agoodman@cfa.harvard.edu}
 
\begin{abstract}
While observational evidence for the episodic nature of young stellar outflows
continues to mount, existing numerical and theoretical models of
molecular outflows assume they are formed by the interaction of a non-episodic
wind from a young stellar object with an ambient cloud. In this {\it Letter} we
estimate and discuss the effects of episodicity on the mass-velocity and
position-velocity relations observed in molecular outflows.  We
explain how many recent
observational results disagree with the predictions of non-episodic 
outflow models,
and we offer simple explanations for the discrepancies. In
particular, we discuss
how an episodic stellar outflow can steepen the power-law slope of the
mass-velocity relation in a molecular outflow. And, we
illustrate how
an episodic outflow can produce multiple ``Hubble-wedges'' in the
position-velocity
distribution of a molecular outflow.  With a little more information 
than we have
now, it may be possible to use the ``fossil record" embedded in
a molecular outflow's mass-velocity and position-velocity relations 
to reconstruct
the history of a young stellar object's mass ejection episodes.
\end{abstract}

\section{Introduction}

As forming stars accrete mass inside molecular clouds, they
simultaneously throw away a large fraction of that mass in a vigorous
outflow. So-called ``molecular" or ``CO" outflows are one of several
manifestations of
these mass outflows from young stellar objects (YSOs). The molecular
outflows consist
primarily of swept-up ambient cloud material, and are most likely driven by  an
underlying (faster) flow emanating from the forming star and/or its
circumstellar
disk.

Spectral-line observations of molecular outflows have shown that they exhibit
several common observational characteristics.  A log-log plot of outflow
mass as a function of velocity (a.k.a.~a ``mass spectrum"), usually shows
power-law mass-velocity ($M-v$) relations where $dM(v)/dv~\alpha~v^{-\gamma}$.
Position-velocity relations of outflows often exhibit another common 
characteristic ---a
so-called ``Hubble-law" where the maximum velocity of the outflowing gas
increases nearly linearly with distance from the source.

Many observational and theoretical studies have found that molecular outflows
have a mass spectrum with $\gamma \sim 2$ (e.g., Lada \& Fich 1996; 
Smith, Suttner, \& Yorke~1997; Matzner \& McKee~1999), and that the outflows can be characterized by a single
Hubble-law (e.g., Meyers-Rice \& Lada~1991, Shu et al.~1991; Lada \& Fich
1996; Matzner \& McKee~1999). But, recent outflow studies
(e.g., Yu, Billawala, \& Bally~1999; Arce \& Goodman~2001)
have shown  $\gamma$ to be much steeper than 2 in some cases.  Furthermore, the
position-velocity diagrams of those steep-mass-spectrum molecular outflows are
characterized by several Hubble-law ``wedges," rather than a single one. In
this {\it Letter}, we show how an {\it episodic}\footnote{ 
We use the term ``episodic outflow'' to mean an outflow that varies in 
shape, mass loss rate, direction, and/or velocity in an unpredictable
way and it is ``not-strictly-periodic''. Episodic outflows (with ages about
$10^4$ to $10^5$ yr) are characterized by events of violent mass ejection 
($\dot{M}_{out} \sim 10^{-5}$ to $10^{-4}$ M$_{\sun}$ yr$^{-1}$) 
every 500 to $10^3$ yr or so, with states of no, or very low 
($\dot{M}_{out} \sim 10^{-8}$ to $10^{-7}$ M$_{\sun}$ yr$^{-1}$) 
mass ejection between the violent
outburst episodes. Several mechanisms which produce episodic outflows from
young stars have been discussed in the literature (e.g., self-regulated thermal
disk instabilities, Bell \& Lin 1994; companion-disk interactions in a multiple 
stellar system, Reipurth 2000, and references therein).} 
outflow could easily 
produce mass
spectra with $\gamma > 2$ and multiple Hubble wedges.

\section{Mass-Velocity Relation}
\subsection{Observing the $M-v$ Relation}

Observers measure molecular outflow mass as a function of velocity 
using molecular
spectral line maps, usually of the \cooz line.  In practice, the mass-velocity
relation is obtained by calculating mass per $\delta v$-wide
velocity bin and plotting mass as a function of velocity offset from 
the host cloud's
mean velocity.  The observed mass-velocity relations to date always imply
more mass at low outflow velocities than at higher outflow 
velocities.  Usually,  $\log (dM/dv)$ vs.~$\log (v)$
has a broken-power-law appearance, with
$dM(v)/dv~\alpha ~v^{-\gamma_{low}}$ for ``low" outflow velocities, and 
$dM(v)/dv~\alpha ~v^{-\gamma_{high}}$
for ``high" velocities.
 The border between ``high" and ``low" is set by the break in 
the power law, and
$\gamma_{high} >
\gamma_{low}$ (see Bachiller \& Tafalla 1999 and references therein).
Hereafter $\gamma$ is used to mean $\gamma_{low}$
whenever $\log (dM/dv)$ has a broken power-law appearance.

Most observational outflow studies estimate outflow mass by
assuming either that \twco emission is optically thin at all outflow
velocities, or by correcting for a constant (velocity-independent) \twco line
opacity. 
In the majority of studies where the outflow mass is obtained in this 
conventional way, $\gamma$ 
ranges from about 0.5 to 3.5, with a concentration
of values near $\gamma \sim 2$ (e.g., Richer et al. 2000).
In fact, however, the \twco line
is often optically thick, and the exact value of opacity depends on velocity.
Not correcting for the velocity-dependent
opacity will, in most cases, underestimate the value of $\gamma$.
A few very recent outflow studies have explicitly
taken the velocity-dependent opacity of the \twco line into
account when estimating outflow mass, and these studies generally 
find $\gamma > 2$
(Bally et al.~1999; Yu et al.~1999; Yu et al.~2000; Arce \& Goodman 2001). 
Discussions on the possible factors that may produce a large value of $\gamma$
can be found in Yu et al.~(1999) and Arce \& Goodman (2001). But,
just because these studies have yielded a steeper mass-spectrum 
slope, on average,
than the ``conventional'' outflow studies, does not necessarily mean that
$\gamma$ would dramatically increase for all
outflows if they were corrected for the velocity-dependent opacity of
$^{12}$CO.
In some cases, usually involving very fast outflows, the assumption that the
\twco is optically thin at the outflow velocities
is fine, and there is no need for opacity correction.
In general, though, we expect that a compilation of
velocity-dependent-opacity-corrected mass spectra would yield a distribution
of $\gamma$'s, with mean $\gamma > 2$. 
As discussed in \S 2.3 below, episodic outflows could produce 
such a distribution
of observed $\gamma$'s.

\subsection{Modeling the $M-v$ Relation}
Numerical and theoretical studies predict that a molecular outflow 
mass spectrum
should have a power-law dependence. In the numerical models of Zhang 
\& Zheng (1997)
and of Smith et al.~(1997), molecular outflows are produced by the
entrainment of ambient gas by a bow shock, and $\gamma$ is found to be
$\sim 1.8$ over most of the outflow velocity range.  Both models also find a
turn-down in the mass spectrum at very high velocities, where $\gamma >>1.8$.
In their numerical models, Downes \& Ray (1999) find that the mass 
spectrum has a
single slope which ranges between 1.58 and 3.75, depending on certain 
parameters of the simulation (i.e., molecular fraction in the jet, amplitude
of the jet velocity variation,\footnote{The models of Smith et al.~(1997) and
Downes \& Ray~(1999) include the effects of rapidly periodic pulsed jets, but
these pulsations are too rapid and regular to be ``episodic'' in our 
definition.} and the ratio of jet to ambient density).
Both Smith et al.~(1997) and Downes \& Ray~(1999)
find that $\gamma$ increases as the jet ages.  According to those models,
the ambient material that once was accelerated by bow shock 
entrainment slows down
as times goes by, which steepens the mass spectrum by shifting mass from the
``fast" end of the mass-velocity relation to the ``slow" end. The 
simulated outflows
of Smith et al.~(1997) and Downes
\& Ray (1999) are relatively young (600 and 300 years, respectively), 
but it seems
reasonable to expect that the qualitative trend of increasing $\gamma$ with age
would continue as the outflow evolves in time, and once-fast gas 
slows down.
Very recent bow
shock modeling by Lee et al. (2000) finds that $\gamma$ also depends 
sensitively on
the inclination of the outflow axis to the line of sight.  

In contrast to the numerical results, the recent analytical study by Matzner \&
McKee (1999; hereafter MM) concludes that it is hard to obtain a value of the
power-law mass spectrum slope that is not very close to 2, for all times.
 Unlike the numerical studies listed
above, however, the entrainment of the ambient medium in
MM is not specifically caused
by a bow shock; rather, it is caused by a collimated wind which sweeps the
ambient gas into a momentum-conserving shell, following the models of Shu et
al.~(1991) and Li \& Shu (1996).   

\subsection {The $M-v$ Relation in Episodic Outflows}

As a framework for a future analytic model of an episodic outflow, consider a
flow as the sum of many collimated MM-like outflows, each of which originates
at a different time,  with a slightly different orientation.  In
this hypothetical picture, the episodic ``bursts" need not all have 
the same total
mass or ejection velocity.  As a result, the bursts can even run over 
themselves in space, with very
fast young bursts overtaking older, slower ones.  In this
picture, an observed molecular outflow represents the {\it time and space
superposition} of many MM-like individual outflow events.  In cases where many
events of comparable energy have taken place within a few dynamical times of
each other, the effects of episodicity will be noticeable in the 
mass-velocity (and
position-velocity, see \S 3) relations measured for the outflow. 
In addition, in order for a burst to effect the molecular outflow, the ejection
angle between the different bursts should be enough so that each travels 
through regions of the cloud that have not yet been cleared out of gas by
previous bursts. Given that most stellar jets have opening angles of a few
degrees (e.g., Mundt, Ray, \& Raga 1991) a change in ejection angle of
only a few degrees between different bursts is enough for a burst to 
easily encounter ``fresh and unperturbed'' cloud material.
In 
the case where
the kinetic energy of one of those events completely dominates any others, the
effects of episodicity will be less noticeable.

MM show that for any stellar wind with a momentum injection rate 
distributed in
space in proportion to $1/\sin^2\theta$, any momentum-conserving interaction
between that wind and an ambient medium with
a power-law density profile will have a mass spectrum with $\gamma \sim
2$.\footnote{Hydromagnetic winds naturally collimate and their resultant force
distribution is proportional to $1/\sin^2\theta$ (see MM and 
references therein).}
So, for the hypothetical episodic flow described above, we can assume that
each outburst of an episodic (hydromagnetic) wind has a $1/\sin^2\theta$
force distribution.  Each outburst of outflowing material will not
necessarily interact with the same ambient gas conditions, as
each successive burst can change the density, velocity and/or temperature
distribution of the ambient gas surrounding the wind.  But, MM clearly
explain that a value of $\gamma$ close to 2 is 
obtained {\it independent} of the  density, velocity
and temperature distributions in the ambient gas.  So, in an episodic
outflow, we can assume that each independent outburst is responsible for putting
ambient gas into motion with a power-law mass-velocity relationship 
with $\gamma \sim 2$.

So, what will the {\it cumulative} observed $\gamma$ for an episodic 
outflow be?
MM say that ``$\gamma \sim 2$" regardless of the time history
of the flow, but that is for a ``time history" where the same flow
is just turned on and off, flowing into an ambient medium with a 
power-law density
distribution.  If instead we envision the bursty flow described above, where
variable amounts of ambient material are accelerated to assorted 
maximum velocities
(due either to variations in the wind, or in the ambient medium), the 
observed mass
spectrum will be the result of superimposed bursts, and will steepen 
(see Figure 1).

Figure~1 shows a mass spectrum for a hypothetical episodic outflow.  Each
molecular outflow episode is represented as a power-law going back to 
zero-velocity\footnote{This is for illustrative purposes only, as 
in reality $\log (dM/dv)$ should flatten at very low velocities.}
with $\gamma = 2$, but with a unique total mass and maximum velocity. 
The observed
molecular outflow mass spectrum is the superposition (sum) of all 
past episodes'
mass spectra.  A power-law fit to the sample ``observed'' mass 
spectrum in Figure
1 results in an ``observed''
$\gamma$ of 2.7.  Thus, an episodic wind composed of a series of MM-like
($\gamma=2$) outbursts is able to create a molecular outflow with $\gamma \geq
2$.  In fact, without additional constraints on the maximum velocity or total
mass of any outburst, practically any value of $\gamma \geq 2$  can be 
created in
this way.

\section{Position-Velocity Relation}
\subsection{Observing {\it p(v)}}

One of the best kinematic diagnostics of outflows
is a position-velocity ({\it p-v}) diagram.  The typical molecular 
outflow {\it p-v}
diagram is constructed by summing spectra in strips perpendicular to 
the assumed outflow axis,
and then contouring line intensity as a function of position along 
the outflow axis and velocity (see
Figure 2).

The simplest molecular outflow {\it p-v} diagrams show maximum 
outflow velocity increasing
approximately linearly as a function of  distance from the source. 
Clear examples of this kind
of ``Hubble-law" velocity distribution are seen in Mon~R2 (Meyers-Rice \& 
Lada~1991) and NGC~2264G
(Lada \& Fich~1996; 
also see bottom panels of Figure 2).  In other cases, {\it p-v}
diagrams exhibit more complicated behavior, which can often be 
described as a series of
``Hubble-wedges" distributed at various distances from the source 
(see top panels of Figure
2).  These multiple wedges are seen in small molecular outflows 
(0.4~pc long) from Class 0
sources (e.g., L~1448, Bachiller et al.~1990)
 as well as in large ($\sim 1$~pc long) 
outflows from Class I sources
(e.g., L1551, Bachiller, Tafalla, \& Cernicharo~1994;
B5 IRS1, Yu et al.~(1999); HH~300, Arce \& Goodman~2001).

\subsection {Modeling {\it p(v)}}

Outflow models can easily produce a {\it single} Hubble-law 
position-velocity relation, but so
far, we do not know of any model that produces multiple Hubble-wedges.

The numerical simulations of Smith et al.~(1997), Zhang
\& Zheng~(1997), and  Downes \& Ray~(1999) in which
outflows are created by the entrainment of ambient
gas by a single bow shock, all
reproduce the Hubble velocity law. In these models, the Hubble velocity
law is in part due to the geometry of the bow shock. A simplified
explanation
is that the highest forward (in the traveling direction of the jet)
velocities  are
found in the apex of the bow shock, which is also the point furthest away
from
the source. The forward velocities decrease toward the wings of the bow
shock,
and the farthest away from the apex (or closer to the outflow source) the
slower
will be the entrained gas (see Figure~3 of Masson \& Chernin~1993; and
Figure~12 of Lada \& Fich~1996).

Bow shock models are not the only ones
that produce the Hubble velocity law; the studies of
Shu et al.~(1991), Li \& Shu~(1996) and MM also obtain molecular outflows
with Hubble velocity laws as a natural consequence of their being created by
a momentum-conserving wind.  This behavior is easy to understand as
``velocity-sorting," since faster material {\it ejected contemporaneously} with
slower material will travel farther from the source.

\subsection {{\it p(v)} in Episodic Outflows}

An episodic stellar wind interacting with ambient gas through bow shock
prompt entrainment could produce a {\it p-v} diagram with several Hubble
wedges.  In an episodic flow with a significantly varying mass-ejection
rate, outflowing gas would show a Hubble-like velocity distribution for each
mass-ejection episode, with the tip of each Hubble wedge at a source offset
corresponding to a bow-shock apex in the outflow map.  Since not all outbursts
from an episodic source necessarily have the same angle with respect 
to the plane
of the sky or accelerate the ambient gas to the same maximum velocity, each
``Hubble wedge'' will not necessarily have the same maximum radial velocity, and
the maximum velocity can change over time.

If a wind either has had only a single outburst or is dominated by one
particularly strong outburst, a molecular outflow with a {\it single} Hubble
velocity law can result (see Figure 2).  Outflow evolution may also be
responsible for creating a single Hubble law.  A molecular outflow from an
episodic wind which had its last outburst (or has not had an outburst in a
considerable amount of time) will be subject to the same kind of
velocity-sorting mentioned above,
and would produce a single Hubble law in its {\it p-v} diagram.

\section{Discussion and Conclusions}

Taken together, $M-v$ relations with $\gamma >2$ and {\it p-v} 
relations showing multiple Hubble
wedges provide very strong evidence that ``episodicity" is a key 
feature of molecular outflows
from young stars.
\subsection {Reconstructing History}
It is tempting to try to reconstruct the outburst history of an episodic
outflow by  combining information from $M-v$ and {\it p-v} diagrams. 
In very simple cases,
where two or three outbursts dominate the flow, this is possible
(e.g., Gueth, Guilloteau, \& Bachiller 1996; Narayanan \& Walker 1996).
In perhaps more typical cases
though, where important  bursts are separated by much less than the 
dynamical time scale for
each burst, modeling is considerably more complicated ---in that 
uniqueness problems plague the
search for solutions.  We expect though, that if an {\it a priori} 
model of the underlying jet
can be made (based on either the molecular line map or other data), 
the consistency of this
model with both the $M-v$ and {\it p-v} relations can be tested.

\subsection {The need for new models}
Almost since the day they were discovered, molecular outflows have 
been thought of as a
possible (re-)supplier of the turbulent energy needed to support 
molecular clouds against rapid
gravitational collapse (e.g. Norman \& Silk 1980 and citations 
thereto).  It has also been
suggested that a molecular outflow may be able to disrupt its host 
core  (e.g, Tafalla \& Myers
1998), and/or that a combination of outflows may  ``churn'' (Bally et 
al.~1999) or disrupt
(Knee \& Sandell 2000) a whole molecular  cloud complex.

Now, dozens of observational papers, only some of which are mentioned 
above, have pointed to
the ``episodic" nature of young stellar outflows.   Simply put, 
observers are now {\it sure} that
these outflows originate from a bursty source.  The outflow source may also 
be precessing and/or
moving through the ISM, and each of its bursts may accelerate a {\it 
different} amount of ambient
matter to a {\it different} velocity.  In this {\it Letter}, we have 
shown that the episodic
nature of outflows causes their {\it M-v} and {\it p-v} relations to 
be very different from those
predicted by any existing analytic model.

In order to correctly asses the energetic role of
outflows in a magnetized ISM, it is now necessary to account for the time 
history of their interaction
with their surroundings. 
The episodic nature of outflows does not affect the {\it total} momentum
injected in the ISM by outflows, and thus it does not matter in most of the
ISM feedback theories (e.g., Norman \& Silk 1980, and citations thereto). 
On the other hand,
outflow episodicity may change the effective length and time scales
on which the momentum is injected, which is important for 
a magnetized ISM with MHD turbulence
(e.g., Padoan \& Nordlund 1999, and references therein). We offer this
{\it Letter} as a motivation for would-be modelers.  Any detailed 
realistic model of a molecular
outflow must include its episodic nature.

\acknowledgements

We are grateful to Charles Lada for expert comments on this work as 
it progressed, Chris Matzner (the referee) for his very useful comments
and to the National Science Foundation (AST-9457456 and AST-9721455) for supporting our efforts.

\clearpage

\begin{figure}
\plotone{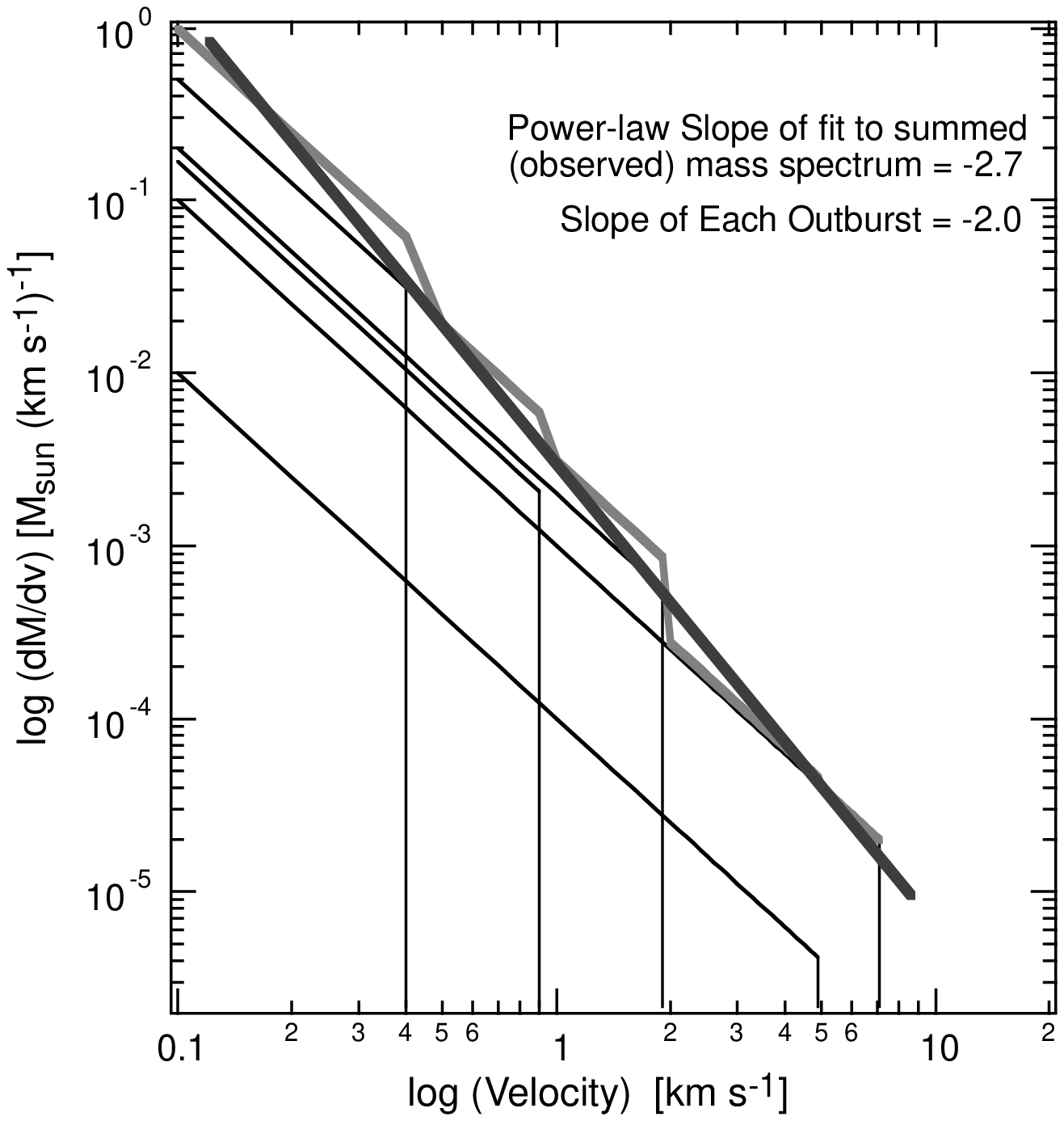}
\caption{Molecular outflow mass spectrum 
for a  hypothetical episodic outflow.
Thin dark lines represent the mass spectrum produced by different 
ejection episodes. The thick grey line is the sum of all the episodes,
and represents the observed mass spectrum. The thick dark line is a fit
to the observed mass spectrum. See \S 2.3.}
\end{figure}

\clearpage

\begin{figure}
\plotfiddle{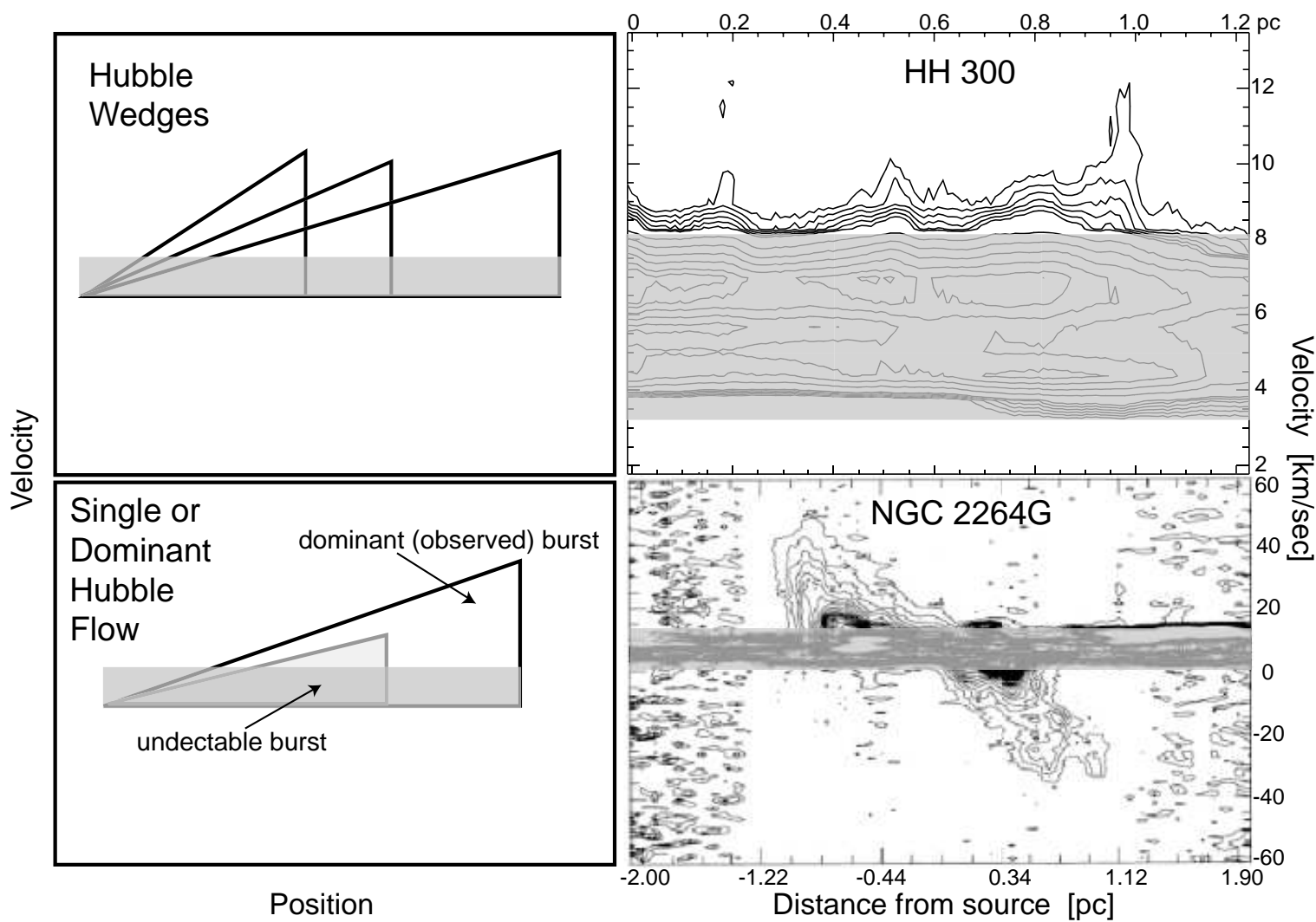}{6in}{-90}{80}{80}{-315}{500}
\caption{{\it Top-Left.} Schematic picture of the 
position-velocity diagram of an episodic outflow. {\it Top-Right.}
Example of an episodic outflow (HH~300), from Arce \& Goodman (2001).
{\it Bottom-Left.} Schematic picture of the position-velocity diagram
of an outflow with only one burst or with one dominant burst. 
{\it Bottom-Right.} Example of a molecular outflow (NGC~2264G) 
with a single Hubble-law velocity distribution (panel based on Figure 4 
of Lada \& Fich [1996]). The grey area represents the ambient 
cloud velocities in which outflow emission cannot be detected, as it is
indistinguishable from the ambient cloud emission.}
\end{figure}

\end{document}